\documentclass[fdp,a4paper,fleqn%
]{w-art}

\usepackage{times,cite,w-thm}

\theoremstyle{plain}

\theoremstyle{definition}

\usepackage[]{graphicx}
\usepackage{amsmath,amsfonts,amssymb,bbm}

\newcommand{\im}{\,\mathbb{I}\mbox{m}\,}

\newcommand{\parfrac}[2]{\frac{\partial #1}{\partial #2}}

\newcommand{\be}{\begin{equation}} \newcommand{\ee}{\end{equation}}
\newcommand{\bea}{\begin{equation} \begin{aligned}} \newcommand{\eea}{\end{aligned} \end{equation}}

\newcommand{\cD}{\mathcal{D}}

\newcommand{\cL}{\mathcal{L}}

\newcommand{\cN}{\mathcal{N}}
\newcommand{\cO}{\mathcal{O}}

\newcommand{\bN}{\mathbb{N}}

\newcommand{\bR}{\mathbb{R}}
\newcommand{\bZ}{\mathbb{Z}}

\def\sl{\mathfrak{sl}}

\def\repa{\raise4pt\hbox{$\square$}\mkern-14mu\raise-4pt\hbox{$\square$}}
\def\repab{\overline{\raise4pt\hbox{$\square$}\mkern-14mu\raise-4pt\hbox{$\square$}\mkern-1mu}}

\def\smileface{\ensuremath{\hbox{\large$\bigcirc$}\mkern-15mu\raise-1pt\hbox{\scriptsize$\smallsmile$}%
\mkern-10mu\raise4pt\hbox{..}\mkern4mu}}
\def\frownface{\ensuremath{\hbox{\large$\bigcirc$}\mkern-15mu\raise-1pt\hbox{\scriptsize$\smallfrown$}%
\mkern-10mu\raise4pt\hbox{..}\mkern4mu}}

\DeclareMathOperator{\Tr}{Tr}

\begin{document}
\DOIsuffix{theDOIsuffix}
\pagespan{1}{}



\title[Holography and condensed matter]{Holography and condensed matter}


\author[F. Benini]{Francesco Benini
  \footnote{Corresponding author\quad E-mail:~\textsf{fbenini@scgp.stonybrook.edu},
            Phone: +1\,631\,632\,2813}}
\address
{Simons Center for Geometry and Physics, State University of New York, \\
Stony Brook, NY 11794-3636, USA}
\begin{abstract}
This is a short review of recent developments in the application of AdS/CFT methods to some condensed matter problems. In particular we present the holographic description of a local quantum critical state, related to (non)-Fermi liquids and the strange metal, that appears in large $N$ CFTs with a gravity dual at finite density and zero temperature, and explore its properties by probing it with fermionic operators. Further we discuss possible bosonic and fermionic instabilities, leading to s-, p- or d-wave holographic superconductors and ``electron stars''. Finally we present a realization of local quantum criticality via an impurity problem.
\end{abstract}
\maketitle                   






\section{Introduction}

Two broad questions often occur in problems related to condensed matter systems: What sort of gapless phases can arise from finite density or charge density states? What kind of physics can emerge at a quantum critical point? In the absence of sharp quasiparticles or weakly coupled effective degrees of freedom in the infrared (IR), answering to these questions can be a hard task and conventional field theory methods might fail. In the last decade a new powerful tool to tackle strongly coupled field theories has been developed: AdS/CFT (alias gauge/gravity correspondence alias holography) \cite{Maldacena:1997re, Gubser:1998bc, Witten:1998qj} (see also the review \cite{Aharony:1999ti}). It is natural to wonder what such method can say about the aforementioned problems. Other reviews, besides the present one, that explore at length the issue are \cite{Hartnoll:2009sz, Herzog:2009xv, McGreevy:2009xe, Horowitz:2010gk, Hartnoll:2011fn}. Here we will focus on the low-temperature limit of simple holographic models, on some models of non-Fermi liquids and on a relation to impurity models.

AdS/CFT is a correspondence between a conformal (non-gravitational) quantum field theory in $d$ dimen\-sions---that we will call the ``boundary theory''---and a quantum theory of gravity on a $d+1$-dimensional background which is asymptotically AdS$_{d+1}$---called the ``bulk theory'' (we refer the reader to the literature \cite{Maldacena:1997re, Gubser:1998bc, Witten:1998qj, Aharony:1999ti}). The fact that the correspondence, strictly speaking, only applies to conformal boundary theories is not a harmful limitation because on the one hand in condensed matter problems we are often interested in the IR physics around a (quantum) critical point regardless of its UV completion, and on the other hand we will consider systems with finite charge density and possibly non-vanishing order parameters where the UV conformality of the boundary theory is anyway broken (while a new scaling symmetry might emerge in the IR).

To each operator of the boundary theory corresponds a field in the bulk. For instance the stress tensor $T_{\mu\nu}$ corresponds to the graviton $g_{\mu\nu}$, a conserved current $J_\mu$ corresponds to a gauge field $A_\mu$, and a scalar order parameter $\Phi$ corresponds to a scalar field $\varphi$. The statement of the correspondence is that the boundary theory partition function $Z[J]$, as a function of sources $J$ coupled to the operators $\cO$ through the action $S_\cO = \int d^dx\, J\cO$, equals the bulk partition function with the boundary condition that fields asymptote to the sources. For instance, for a scalar operator $\Phi$ of dimension $\Delta$ the boundary condition around $r \to 0$ is:
\be
\varphi \,\sim\, J \, r^{d-\Delta} + \langle \Phi \rangle\, r^\Delta \;,
\ee
where $r$ is a ``radial'' coordinate that goes to zero at the boundary and is positive in the interior of AdS$_{d+1}$. The vacuum expectation value (VEV) $\langle \Phi \rangle$ is not part of the boundary conditions, but rather is read off from the gravity solution. The partition function of the bulk quantum theory of gravity can be very difficult to define. However in full-fledged example the boundary theory is usually a non-Abelian gauge theory with $N$ colors, and in a large $N$ and large 't Hooft coupling limit the gravity theory becomes classical. In such limit the correspondence states that
\be
Z_\text{boundary}[J] = e^{iS_\text{bulk}[\varphi_0]} \Big|_{\varphi_0 \,\sim\, J\, r^{d-\Delta} + \langle \Phi \rangle\, r^\Delta}
\ee
where $\varphi_0$ is a classical solution of the bulk equations of motion (EOMs) subject to the boundary conditions.

The correspondence can be extended to boundary theories at finite temperature $T$: in this case the gravity solution contains a black hole (BH) whose horizon has temperature $T$. Extra regularity boundary conditions have to be imposed at the horizon, depending on the signature and the type of correlators one is interested in. On the other hand we are interested in finite (charge) density states. The easiest way to obtain a finite charge density state is to start with a theory with a symmetry, say a $U(1)$ symmetry for simplicity, and introduce a chemical potential $\mu$, that is a source term $\mu J^0$ into the Lagrangian. This is achieved by imposing the corresponding boundary condition to the gauge field $A_0$.

\subsection{Emergent gauge fields}
\label{sec: emergent gauge}

In most condensed matter problems the only gauge field present is the photon, a $U(1)$ gauge field. On the contrary, AdS/CFT becomes most computable when the boundary theory is in the large $N$ limit. Indeed AdS/CFT is exploited by considering a CFT with a \emph{global} U(1) symmetry---the photon is thus a spectator that might be weakly gauged at the end of the day---and a large $N$ strongly coupled \emph{gauge} symmetry which keeps the system at strong coupling. Such gauge fields, in relation to the original condensed matter problem, have to be thought of as emergent IR degrees of freedom, not visible outside the fixed point. As part of our motivations, we would like to give one concrete example of emergent gauge field.

The following example has been studied in \cite{Senthil:2004a, Senthil:2004b} and we will follow \cite{Sachdev:2007}. Consider a square lattice spin-half model, with Hamiltonian:
\be
\label{Hamiltonian spin}
H = J \sum_{\langle ij \rangle} \vec S_i \cdot \vec S_j - Q \sum_{\langle ijkl \rangle} \Big( \vec S_i \cdot \vec S_j - \frac14 \Big) \Big( \vec S_k \cdot \vec S_l - \frac14 \Big) \;,
\ee
where $J,Q>0$ are coupling constants and $\vec S_i$ are three-dimensional spins. The first summation is over nearest neighbor sites, whilst the second is over squared plaquettes. The system has a $\bZ_4$ symmetry that rotates the square lattice, and a $spin(3)$ symmetry that rotates the spins.

To begin with, consider the limit $Q/J \to 0$: the system asymptotes to the isotropic Heisemberg antiferromagnet, whose ground state has N\'eel order
\be
\langle \vec S_i \rangle = (-1)^i \vec \Phi \;,
\ee
where we mean that signs are alternating, and breaks the spin rotation symmetry. The low energy excitations are spin density waves of $\vec\Phi$ (spin-triplets), described by a mean-field IR effective Lagrangian. With a change of variables we can represent the vector $\Phi^a$ as a bi-spinor
\be
\Phi^a = z^*_\alpha \sigma^a_{\alpha\beta} z_\beta \;,
\ee
where $z_\alpha$ is a complex spinor and $\sigma^a_{\alpha\beta}$ are gamma-matrices. The new degrees of freedom are redundant, though, as a phase rotation of $z_\alpha$ does not affect $\Phi^a$. We should then gauge such phase away, with the introduction of a $U(1)$ gauge field $A_\mu$. The effective IR Lagrangian is:
\be
\label{Lag emergent gauge}
\cL_{z,A} = - \big| (\partial - iA) z \big|^2 - s |z|^2 - u |z|^4 - \frac1{2e^2} F^2 \;,
\ee
with $u>0$ and $s$ below a critical value $s_c$. Therefore $z_\alpha$ condenses, the gauge field is Higgsed and it might be integrated out: at this stage the introduction of $A_\mu$ looks as an unnecessary formal manipulation.

On the contrary, consider the limit $Q/J \to \infty$. The second term in \eqref{Hamiltonian spin} favors the arrangement of the spins in neighboring spin-singlet pairs---the ground state has valence-bond-solid (VBS) order which breaks the $\bZ_4$ lattice rotation symmetry (while preserving spin rotation symmetry). The order parameter is the operator:
\be
\Psi = (-1)^{j_x} S_j \cdot S_{j+ \hat x} + i(-1)^{j_y} S_j \cdot S_{j+ \hat y}
\ee
where $j_{x,y}$ are the lattice coordinates. The low energy excitations come from breaking a spin-singlet into two free spins $z_\alpha$, and are again described by the Lagrangian \eqref{Lag emergent gauge} with $s$ above a critical value $s_c$, if we identify the $U(1)$ topological symmetry (shift $\zeta \to \zeta + \delta$ of the scalar $\zeta$ dual to $A_\mu$) with the lattice rotation symmetry, and include a term $\cL_\zeta \sim \cos \frac{8\pi \zeta}{e^2}$ in the Lagrangian so to explicitly break such topological $U(1)$ to $\bZ_4$. Such term makes the dual photon $\zeta$ massive. Moreover $\Psi$ can be identified with the monopole operator.

Now \cite{Senthil:2004a, Senthil:2004b} observe that if we tune $s$ to the critical value $s_c$, the extra term $\cL_\zeta$ becomes irrelevant and both $z,A_\mu$ stay massless at the critical point. They provide the effective description of \eqref{Hamiltonian spin} at the critical value of $Q/J$, and $A_\mu$ is an emergent gauge field (not present on either side).

\section{Finite density states and AdS$_2$}
\label{sec: AdS2}

In the following we will focus on $2+1$-dimensional boundary theories, corresponding to a gravity theory in asymptotically AdS$_4$. The minimal setup to describe finite charge density states includes the graviton $g_{\mu\nu}$ and a $U(1)$ gauge field $A_\mu$. We will consider the simple Lagrangian:
\be
\label{Lagrangian basic setup}
\cL = \frac1{2\kappa^2} \Big( R + \frac6{L^2} \Big) - \frac1{4e^2} F_{\mu\nu} F^{\mu\nu} + \dots
\ee
where dots stand for other fields that will play a r\^ole later on. Imposing the boundary conditions for chemical potential $\mu$ and temperature $T$, the solution is the Reissner-Nordstr\"om-AdS black hole:
\be
\label{RNAdS BH}
ds^2 = \frac{L^2}{r^2} \Big( -f(r) dt^2 + \frac{dr^2}{f(r)} + dx^2 + dy^2 \Big) \;,\qquad A = \mu \Big( 1 - \frac r{r_+} \Big) \, dt
\ee
where $r$ is a ``radial'' coordinate that vanishes at the boundary and is positive in the interior, $f(r)$ is a blackening function and $r_+$ is the position of a horizon:
\be
f(r) = 1 - \Big( 1 + \frac{r_+^2 \mu^2}{2\gamma^2} \Big) \Big( \frac r{r_+} \Big)^3 + \frac{r_+^2 \mu^2}{2\gamma^2} \Big( \frac r{r_+} \Big)^4 \;,\qquad T = \frac1{4\pi r_+} \Big( 3 - \frac{r_+^2 \mu^2}{2\gamma^2} \Big) \;.
\ee
We introduced the parameter $\gamma \equiv eL/\kappa$.

Taking the zero-temperature and near-horizon limit, the geometry asymptotes to $AdS_2 \times \bR^2$:
\be
\label{AdS2 vacuum}
ds^2 \,\to\, \frac{L^2}6 \Big( \frac{-dt^2 + dr^2}{r^2} \Big) + dx^2 + dy^2 \;,\qquad A = \frac{\gamma}{\sqrt6}\, \frac{dt}r \;,
\ee
in terms of new coordinates $r,t,x,y$.
Surprisingly, in the IR the state has an emergent ``local'' scaling symmetry: $r\to \lambda r$, $t \to \lambda t$, $x,y \to x,y$. We can think of it as the $z \to \infty$ limit of a Lifshitz scaling (sec. \ref{sec: Lifshitz}). As we will see later in section \ref{sec: impurity}, a possible lattice realization is through an impurity model. Such state has some peculiar properties. First, the entropy density (computed from the black hole horizon area $A$ as $s = 2\pi A/\kappa^2 vol$) has a non-vanishing zero-temperature limit:
\be
s(T \to 0) = \frac{\pi \mu^2}{3e^2} \;.
\ee
Second, the density of states is IR divergent \cite{Jensen:2011su}: $\rho(E) \sim e^s \delta(E) + E^{-1}$. One thus expects some instability to kick in somewhere in the IR. We will later consider two (non-exhaustive) possibilities: Bose-Einstein condensation and population of a Fermi sea.

\subsection{Probe fermions and fermionic spectral functions}
\label{sec: probe fermions}

In order to gather more information about the state \eqref{AdS2 vacuum}, one can compute 2-point functions of probe fermionic operators \cite{Lee:2008xf, Liu:2009dm, Cubrovic:2009ye, Faulkner:2009wj}. A bulk probe Dirac fermion $\psi$ of charge $e$ and mass $m$ corresponds to a fermionic operator $\cO_\psi$ of dimension%
\footnote{In the range $|m|L < \frac12$ an alternative quantization $\Delta = \frac32 - |m|L$ is possible.}
$\Delta = \frac32 + |m|L$. To compute 2-point functions only the quadratic action is needed, and we will consider the Dirac Lagrangian:
\be
\label{Dirac Lagrangian}
\cL_\psi = i\bar\psi \Gamma^\mu \Big( \partial_\mu + \frac14 \omega_\mu^{ab} \Gamma_{ab} - iA_\mu \Big) \psi - im \bar\psi \psi \;.
\ee
For parameters that satisfy $(mL)^2 \leq \gamma^2$, there is Schwinger pair production in the bulk \cite{Pioline:2005pf}, and one should expect a finite density of fermions hoovering outside the charged horizon.

It is particularly interesting to compute the single-particle retarded Green's and spectral functions:
\be
G_R(t,\vec x) \equiv i \Theta(t) \langle \{ \cO_{\psi}(t,\vec x), \cO_\psi^\dag(0) \} \rangle \;,\qquad A(\omega, \vec k) \equiv \frac1\pi \im G_R(\omega, \vec k) \;,
\ee
where $\Theta(t)$ is the Heaviside step function. The spectral function $A$ describes the density of states, and a surface of sharp peaks represents the dispersion relation $\omega(\vec k)$ of quasinormal modes. One can make a connection with photo-emission (ARPES) experiments on different materials, where such density of states is measured \cite{Damascelli:2003}.

It turns out \cite{Faulkner:2009wj, Faulkner:2010zz} that in the IR CFT the operator $\cO_\psi(\vec k)$ at momentum $\vec k$ has dimension $\delta_k$:
\be
\delta_k = \frac12 + \nu_k \;,\qquad \nu_k \equiv \frac1{\sqrt6} \, \sqrt{ m^2L^2 + \frac{3k^2}{\mu^2} - \gamma^2}
\ee
and its Green's function (in the IR CFT) is $\varsigma_k(\omega) = c(k) \, \omega^{2\nu_k}$, where $c(k)$ is some analytic function of $k$.

For $(mL)^2 < \frac23 \gamma^2$ the Dirac equation has static normalizable solutions \cite{Liu:2009dm} (see their figure 2), which signal a Fermi surface at momentum $k_F$. While the precise value of $k_F$ depends on the details of the UV theory, the physics around it does not.

The small-frequency expansion of the Green's function around the Fermi surface fits the following asymptotic expression:
\be
\label{holographic Green's function}
G_R(\omega, k) \simeq \frac{h_1}{k - k_F - \frac1{v_F} \omega - \Sigma(\omega, k)} \;,\qquad \Sigma(\omega,k) = h_2\,  \varsigma_{k_F}(\omega) \;,
\ee
where $h_1, h_2, v_F$ are constants while $\Sigma(\omega,k)$ at leading order does not depend on $k$ (only on $k_F$).
Let us compare such behavior with that in Landau Fermi liquid (LFL) theory:
\be
G_R^\text{(LFL)}(\omega, k) = \frac{Z}{\omega - v_F (k-k_F) + i \Gamma} + \dots \;,\qquad \Gamma \sim \omega^2 \;.
\ee
Following the pole in the lower complex $\omega$-plane as a function of $k$, we deduce the dispersion relation $\omega_c(k)$ that we split into real and imaginary part: $\omega_c(k) = \omega_*(k) - i \Gamma(k)$. The behavior of the holographic liquid \eqref{holographic Green's function} depends on the value of the parameter $\nu_{k_F}$ \cite{Faulkner:2009wj, Faulkner:2010zz}. For $\nu_{k_F} > \frac12$ we have a Fermi liquid, characterized by sharp quasiparticles:
\be
\omega_*(k) = v_F(k-k_F) + \dots \;,\qquad \frac{\Gamma(k)}{\omega_*(k)} \propto (k-k_F)^{2\nu_{k_F} - 1} \to 0 \;,\qquad Z = h_1 v_F \;.
\ee
We observe, respectively, linear dispersion with Fermi velocity $v_F$; decay rate that goes to zero (stable quasiparticles); non-vanishing spectral weight.
The case $\nu_{k_F}=1$, studied in \cite{Cubrovic:2009ye}, is very similar to a Landau Fermi liquid, although corrections of the form%
\footnote{Logarithmic corrections appears for any $\nu_{k_F} \in \bN/2$ \cite{Faulkner:2009wj}.}
$\omega^2 \log \omega$ are present in $\Gamma$ so that it is not a conventional LFL yet.
For $\nu_{k_F}<\frac12$ we have a non-Fermi liquid, with no sharp quasiparticles:
\be
\omega_*(k) \sim (k-k_F)^{1/2\nu_{k_F}} \;,\qquad \frac{\Gamma(k)}{\omega_*} \to \text{const} \;,\qquad Z \propto (k-k_F)^\frac{1-2\nu_{k_F}}{\nu_{k_F}} \to 0 \;.
\ee
In this case we have non-analytic dispersion relation, with imaginary part always comparable with its real part (quasiparticles never stable); vanishing spectral weight.
The boundary case $\nu_{k_F} = \frac12$ is particularly interesting as it provides a realization of the ``marginal Fermi liquid''  (MFL) phenomenological model introduced in \cite{Varma:1989zz}. The small-frequency behavior is
\be
G_R^\text{(MFL)} \simeq \frac{h_1}{(k-k_F) + c_R \omega \log \omega + c_1 \omega} \;,\qquad Z \sim \frac1{|\log \omega_*|} \to 0 \;,
\ee
where $c_1$ is complex while $c_R$ is real. The single-particle scattering rate is suppressed with respect to the real part, but only logarithmically; the quasiparticle residue vanishes, but only logarithmically.

At finite temperature $T \ll \mu$, the Green's function pole never reaches the real axis and the Fermi surface gets smeared. One finds the two asymptotic behaviors \cite{Faulkner:2010zz}:
\be
\omega \ll T: \quad \Sigma(\omega,k) \propto T^{2\nu_k} \;,\qquad\qquad \omega \gg T:\quad \Sigma(\tau,k) \sim \left| \frac{\pi T}{\sin(\pi T \tau)} \right|^{2\Delta_k} \;.
\ee

The leading contribution of the Fermi surface to the conductivity, which can be computed with Kubo's formula $\sigma(\omega) = \frac1{i\omega} \langle J_x(\omega) J_x(-\omega) \rangle_\text{retarded}$, is evaluated \cite{Faulkner:2010zz} by a one-loop diagram in the bulk where the fermions run in the loop connecting two insertions of $A_\mu$. The resulting DC conductivity, for $\nu_k \leq \frac12$, is:
\be
\sigma(\omega \to 0) \sim T^{-2\nu_k} \;.
\ee
In particular, when the parameter $\nu_{k_F}$ is set to $\nu_{k_F} = \frac12$ one gets a contribution to the resistivity linear in temperature. This is precisely the behavior observed in the strange metal phase of cuprates (see \textit{e.g.} \cite{Damascelli:2003}), whose theoretical origin has not been fully understood yet.

\subsection{Semi-holographic (non)-Fermi liquids}
\label{sec: semiholographic}

The IR Green's and spectral functions discussed in the previous section can be reproduced by a simple effective or semi-holographic model \cite{Faulkner:2010tq}. Consider a Fermi liquid $\Psi$ coupled to the fermionic fluctuations $\chi$ of a critical system with large dynamical exponent $z$---for simplicity we will consider a local critical system:
\be
\cL = i \big[\overline{\Psi} (\omega - v_F k_\perp)\Psi + g\overline\Psi \chi + g^*\bar\chi \Psi + \bar\chi \varsigma^{-1} \chi \big] \;,
\ee
where $k_\perp = k - k_F$ while $\varsigma$ is the local critical system Green's function: $\varsigma = \langle \bar\chi \chi \rangle = c(k) \omega^{2\nu_k}$. By resumming the series of interactions \cite{Faulkner:2010tq} one obtains the effective Green's function:
\be
\langle \overline\Psi \Psi \rangle = \frac1{\omega - v_F k_\perp - |g|^2\varsigma} \;,
\ee
which has the same form as the ones discussed before.

\section{Instabilities: superconductors and electron stars}

We want to consider now what effects other fields, represented by dots in the Lagrangian \eqref{Lagrangian basic setup} and generically present in the bulk, can have. The fields in \eqref{Lagrangian basic setup} are the ones responsible for the backgrounds we discussed, while the extra fields will be coupled to them and might become unstable. The two main mechanisms we will discuss are Bose-Einstein condensation (BEC) in the case of bosons, and population of the Fermi sea in the case of fermions.

\subsection{Holographic superconductors}
\label{sec: superconductors}

A first type of instability might arise when an order parameter $\cO_\phi$---charged under the $U(1)$ symmetry with chemical potential---is present. Consider a charged scalar field $\phi$ \cite{Gubser:2008px, Hartnoll:2008vx} with the minimally coupled Lagrangian:
\be
\label{Lagrangian superconductor}
\cL = \frac1{2\kappa^2} \Big( R + \frac6{L^2} \Big) - \frac1{4e^2} F_{\mu\nu} F^{\mu\nu} - \big| \nabla \phi - i A \phi \big|^2 - m^2 |\phi|^2 - V(|\phi|)
\ee
where $V$ is some potential. In the presence of a bulk electric flux (that follows from the boundary chemical potential), one could expect BEC ending up with a background:
\be
\label{superconductor ansatz}
\frac{ds^2}{L^2} = - f(r) dt^2 + g(r) dr^2 + \frac{dx^2 + dy^2}{r^2} \;,\qquad A = \gamma \, h(r) dt \;,\qquad \phi = \phi(r)
\ee
where $f,g,h,\phi$ are radial functions. Whether the condensation takes place depends on which state has lower free energy. Indeed, upon numerical evaluation, it turns out \cite{Hartnoll:2008vx, Hartnoll:2008kx, Gubser:2008pf, Denef:2009tp} that at $T=0$, $\phi$ condenses whenever its effective mass in the IR AdS$_2$ region falls below the AdS$_2$ BF bound \cite{Breitenlohner:1982bm}:
\be
\frac16 \big( m^2L^2 - \gamma^2 \big) \leq - \frac14 \;.
\ee
Indeed this is the bound on parameters to have Schwinger pair production in the bulk \cite{Pioline:2005pf}. The condensation takes place up to some critical temperature $T_c$ of order $\mu$. The macroscopically occupied ground state spontaneously breaks the $U(1)$ symmetry, and realizes a ``holographic'' superfluid (if the symmetry is weakly gauged, then it is a superconductor).

Lots of properties of holographic superfluid states have been studied (we refer the reader to the literature, see also \cite{Horowitz:2008bn, Gubser:2008wz}). One key feature is that the condensate undergoes a mean-field second-order phase transition at the critical temperature:
\be
\cO_\phi/T_c \sim (T - T_c)^{1/2} \;.
\ee
The optical conductivity $\sigma(\omega)$ shows in its real part a gap $\Delta_\omega$. An interesting fact is that the ratio $\Delta_\omega/T_c$ is in these systems around 8 \cite{Hartnoll:2008vx}, at least in a limit of large charge $e$. Such number compares well with what measured in some cuprate high-$T_c$ superconductors, while in BCS theory it takes a value around $3.5$.

\subsection{Fermions in superconductors}
\label{sec: fermions superconductors}

Once a superfluid/superconducting state has been established, one can proceed as in section \ref{sec: probe fermions} to study its properties by analyzing two-point functions of fermionic operators $\cO_\Psi$ \cite{Faulkner:2009am, Gubser:2009dt}, corresponding to Dirac fermions $\Psi$ in the bulk. In the bulk Lagrangian only terms up to quadratic order in the fermions are relevant to such computation, however it is essential to keep into account couplings to the background. In \cite{Faulkner:2009am} the following Lagrangian has been considered:
\be
\label{Majorana-like Lagrangian}
\cL_\Psi = i \overline\Psi ( \Gamma^\mu D_\mu - m ) \Psi + \eta_5^* \phi^* \overline{\Psi^c} \Gamma_5 \Psi + \text{h.c.} \;,
\ee
which requires the fermions to have half the charge of $\phi$, and where $\Psi^c$ is the charge conjugate to $\Psi$.

The Majorana-like coupling is crucial because it leads to a gapping of the Fermi surface. It couples positive- to negative-frequency modes, as in a BCS s-wave superconductor. When computing the Green's function $G_R(\omega, \vec k)$, one should look for solutions to the Dirac equation from \eqref{Majorana-like Lagrangian} of the form:
\be
\Psi = e^{-i\omega t + i \vec k \cdot \vec x} \Psi^{(\omega, \vec k)}(r) + e^{i\omega t - i \vec k \cdot \vec x} \Psi^{(-\omega, - \vec k)}(r)
\ee
(the details on how to compute $G_R$ can be found in \cite{Iqbal:2009fd}).
Without the Majorana-like coupling ($\eta_5 = 0$), solutions to the Dirac equation that satisfy the boundary conditions determine quasinormal modes of $\Psi$ whose dispersion relation would be $ \omega = \omega_*(k)$---and its intersection with the plane $\omega = 0$ would determine the Fermi surface at $k_F$. The dispersion relation for the quasinormal modes of $\Psi^c$ would be $\omega = - \omega_*(-k)$. Once $\eta_5$ is turned on, the quasinormal modes of $\Psi$ and $\Psi^c$ are coupled: they cross at $\omega = 0$ and eigenvalue repulsion determines a gapping of the would-be Fermi surface \cite{Faulkner:2009am}.

\subsection{p-wave superconductors}

p-wave superfluids/superconductors are characterized by a spin-1 order parameter. Since the order parameter $\vec W$ is complex, different patterns of symmetry breaking are possible: in a state with $p$ order spatial rotations are broken, while in a $p+ip$ state time reversal is broken while spatial rotations are preserved (up to charge rotations).
Examples are Sr$_2$RuO$_4$ (whose ground state has $p+ip$ order) and $^3$He (which is $p+ip$ at ambient pressure, and $p$ at high pressure). A holographic model \cite{Gubser:2008wv, Ammon:2009xh} can be obtained from a CFT with non-Abelian symmetry, say $SU(2)$, explicitly broken by a $U(1)$ chemical potential. The bulk Lagrangian is
\be
\cL_p = \frac1{2\kappa^2} \Big( R + \frac6{L^2} \Big) - \frac1{4g^2} F_{\mu\nu}^a F^{\mu\nu}_a \;.
\ee
A chemical potential along $\tau^3$ in color space breaks the $SU(2)$ symmetry to $U(1)$ and generates a charged massive W-boson, which becomes the possibly condensing spin-1 order parameter. One can consider a $p$ and a $p+ip$ ansatz respectively:
\be
A_{(p)} = \Phi(r) \tau^3 dt + w(r) \tau^1 dx \;,\qquad A_{(p+ip)} = \Phi(r) \tau^3 dt + w(r) (\tau^1 dx + i\tau^2 dy) \;.
\ee
At large $g$, $p+ip$ is unstable to decay to $p$ \cite{Gubser:2008wv}, but it is not known what happens at small $g$.

\begin{figure}
\begin{minipage}{0.37\textwidth}
\includegraphics[width=\textwidth]{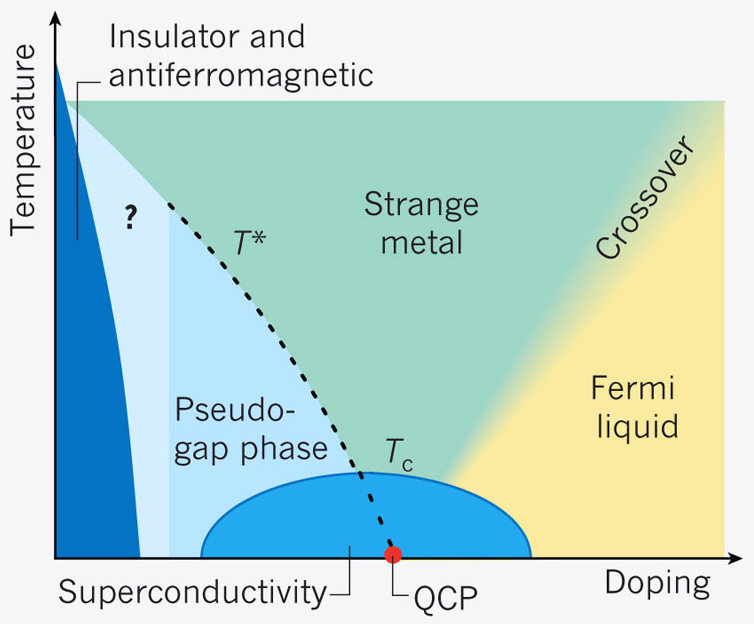}
\caption{Schematic phase diagram of the cuprates showing temperature versus hole doping. Below the curve $T^*$ a pseudo-gap with Fermi arcs opens in the quasiparticle spectrum. Image taken from \cite{Varma:2010}.
\label{fig: cuprate phase diagram}}
\end{minipage}
\hfil
\begin{minipage}{0.60\textwidth}
\includegraphics[width=\textwidth]{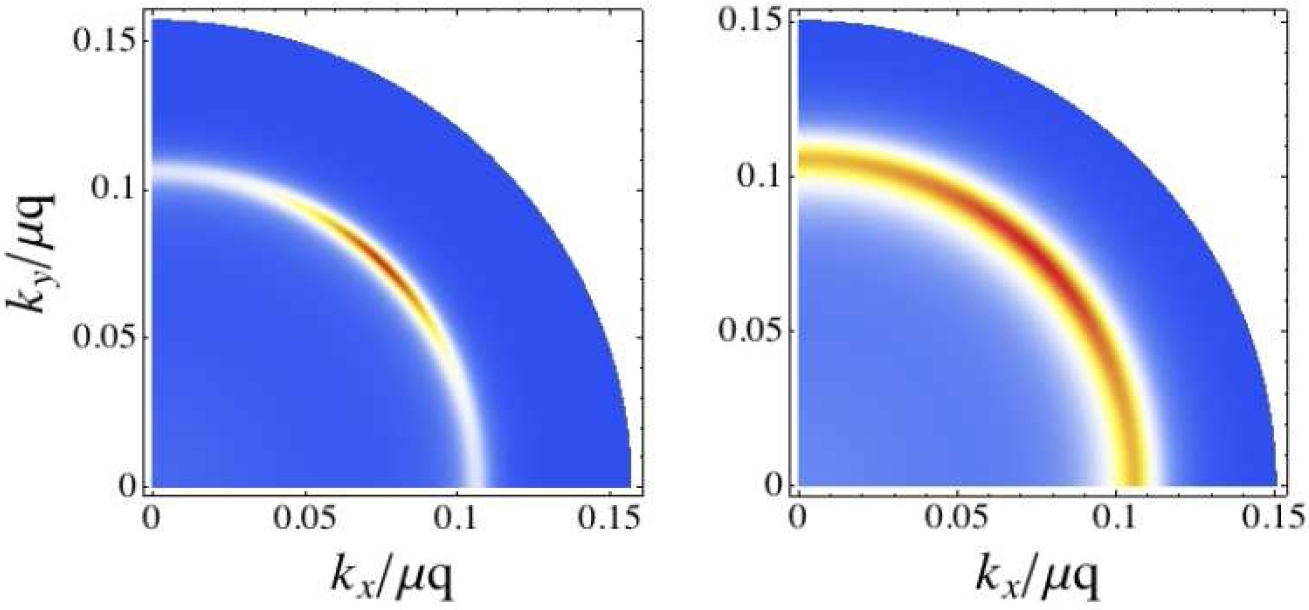}
\caption{Density plot of the fermion spectral function evaluated at $\omega = 0$ for temperatures $T = 0.49T_c$ (left) and $T=0.59 T_c$ (right). Red and blue correspond to large and small values of the spectral function. Image taken from \cite{Benini:2010qc}.
\label{fig: spectral function}}
\end{minipage}
\end{figure}

\subsection{d-wave superconductors}

d-wave superfluids/superconductors have a spin-2 order parameter. Examples are the high-temperature cuprate superconductors. They present a particularly rich phenomenology and phase diagram (fig. \ref{fig: cuprate phase diagram}), also related to the nature of the order parameter. For a review see \cite{Damascelli:2003}.
For instance, in the superconductive phase the Fermi surface is gapped  in an anisotropic way:%
\footnote{Cuprates have a layered molecular structure, so that the physics is mainly $2+1$-dimensional.}
the dependence of the gap $\Delta_\omega$ on the direction in momentum space is $\Delta_\omega \propto |\cos 2\theta|$ (where $\theta$ is an angle). Along the would-be Fermi surface there are four ``Dirac nodes'' where the gap vanishes, and the dispersion relation of quasinormal modes takes the shape of anisotropic Dirac cones. Above the superconductive dome in the underdoped region there is a ``pseudo-gap phase'' where the nodes open up into Fermi arcs whose (angular) length is linear in temperature. One could wonder whether a holographic model would possess such properties.

To reproduce a d-wave order parameter, a holographic model must contain a massive charged spin-2 field in the bulk \cite{Benini:2010pr}. Writing down a consistent and causal action for such field is knowingly hard \cite{Velo:1972, Buchbinder:2000fy}. One possibility would be to take a model that admits a background with compact directions, and perform a Kaluza-Klein (KK) reduction: from the graviton one obtains a KK tower of massive charged spin-2 fields. All these fields will presumably condense at roughly the same temperature, thus solving numerically for the background might be challenging. Another possibility \cite{Benini:2010pr} is to work in a limit of large condensate charge $q$ (at fixed $\mu q$), in which the spin-1 and spin-2 fields do not backreact on the metric: then it is possible to write down a consistent Fierz-Pauli-like action. The background is the AdS-Schwarzschild  black hole, while the matter action is:
\begin{multline}
\cL_d = - |D_\rho \varphi_{\mu\nu}|^2 + 2 |\varphi_\mu|^2 + |D_\mu \varphi|^2 - (\varphi^{*\mu} D_\mu \varphi + \text{c.c.}) - m^2 \big( |\varphi_{\mu\nu}|^2 - |\varphi|^2 \big) \\
+ 2 R_{\mu\nu\rho\lambda} \varphi^{*\mu\rho} \varphi^{\nu\lambda} - \frac14 R |\varphi|^2 - i q F_{\mu\nu} \varphi^{*\mu\lambda} \varphi^\nu_\lambda - \frac14 F_{\mu\nu} F^{\mu\nu} \;,
\end{multline}
where $\varphi_{\mu\nu}$ is a symmetric tensor, $\varphi_\mu \equiv D^\nu \varphi_{\nu\mu}$, $\varphi \equiv \varphi^\mu_\mu$ and $D_\mu = \nabla_\mu - iqA_\mu$. The action is ghost-free and leads to a hyperbolic system of EOMs, that is to a well-posed Cauchy problem. On the other hand it leads to superluminality, which could be cured by higher order terms: such corrections are small in a limit of large $mL$.

Both a $d$ and a $d+id$ ansatz\"e are possible. Let us consider a $d$ ansatz: $\varphi_{xx} = - \varphi_{yy} \equiv \varphi_\Delta(r)$, $A = \Phi(r) dt$. It turns out that, after a suitable rescaling, the equations for $\varphi_\Delta, \Phi$ are identical to the s-wave case: in particular there is condensation below a critical temperature $T_c$.

As in sections \ref{sec: probe fermions} and \ref{sec: fermions superconductors}, the superfluid/superconducting state is conveniently probed by 2-point functions of fermionic operators. In \cite{Benini:2010qc} the following Lagrangian, quadratic in the fermions, has been considered:
\be
\label{Lagrangian fermions in d}
\cL_\Psi = i \overline\Psi (\Gamma^\mu D_\mu - m) \Psi + \eta_5^* \varphi_{\mu\nu}^* \overline{\Psi^c} \Gamma^\mu D^\nu \Psi + \text{h.c.} \;.
\ee
If we restrict ourselves to couplings of dimension smaller that six, to a condensate with twice the charge of the fermions (natural if the order parameter is a sort of Cooper pair) and to the $d$ ansatz, the Lagrangian \eqref{Lagrangian fermions in d} is essentially uniquely fixed.%
\footnote{Other two possible terms are $|\varphi_{\mu\nu}|^2 \overline \Psi(c_1 + c_2 \Gamma_5)\Psi$ which corrects the fermion mass, and the dipole term $\overline \Psi \Gamma^{\mu\nu} F_{\mu\nu} \Psi$ which does not depend on the condensate.}

\begin{figure}
\centering
\includegraphics[width=0.9\textwidth]{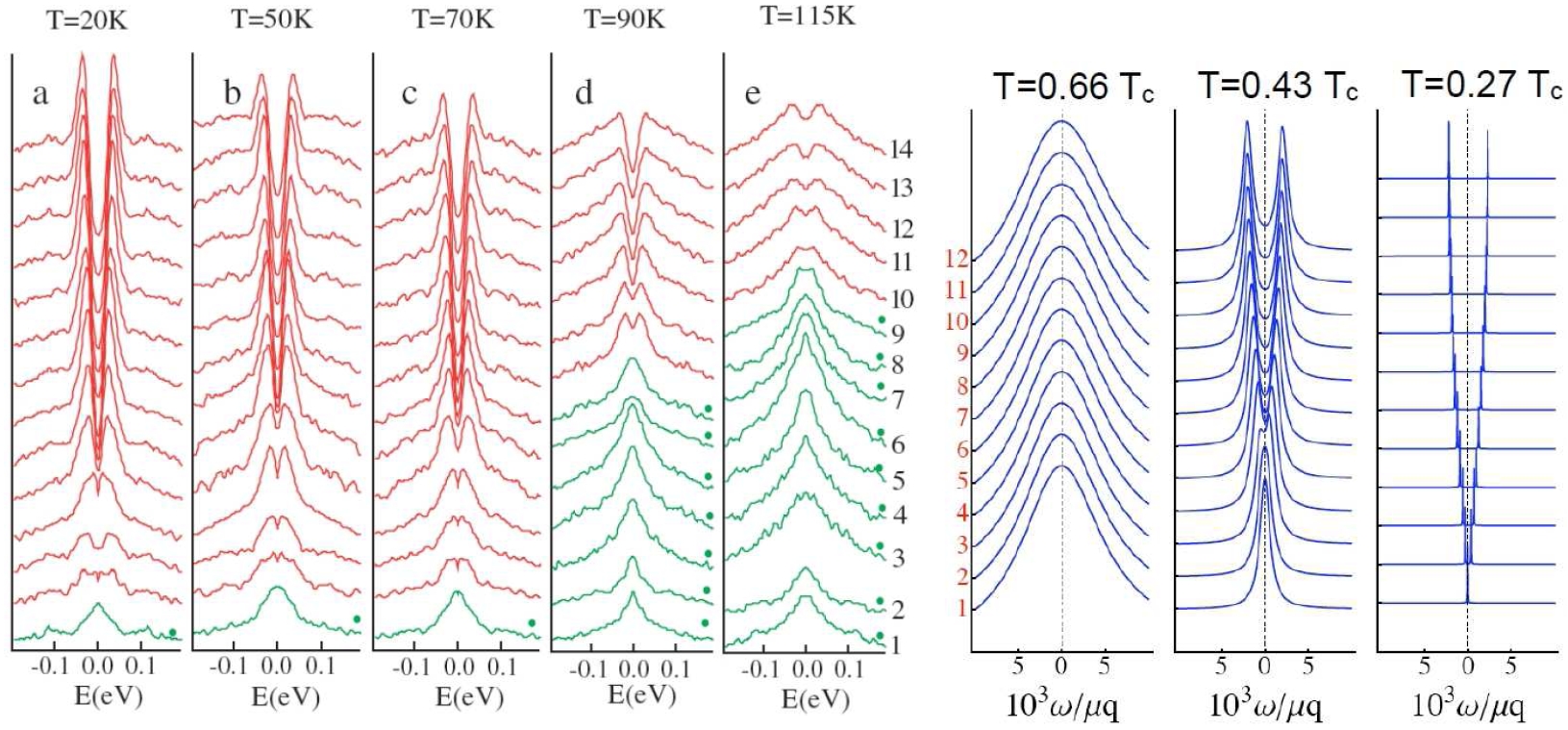}
\caption{EDCs at the would-be Fermi momentum $k_F(\theta)$ for several angles $\theta$ in momentum space and temperatures. Angles run from $0$ to $\frac\pi4$ with homogeneous spacing. Left (from \cite{Kanigel:2007}): underdoped Bi$_2$Sr$_2$CaCuO$_8$; red curves show a gap, green curves show no gap. Right (from \cite{Benini:2010qc}): the holographic model.
\label{fig: EDCs}}
\end{figure}

\begin{figure}
\centering
\includegraphics[width=0.9\textwidth]{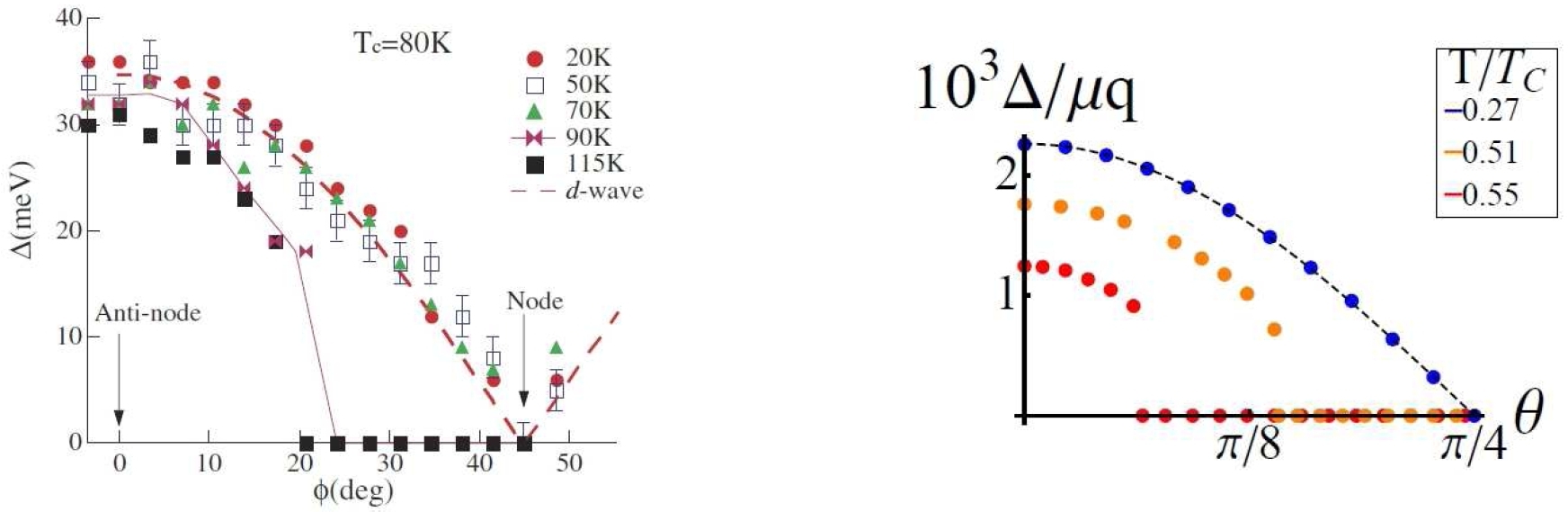}
\caption{Gapping $\Delta_\omega(\theta)$ of the Fermi surface as function of the angle $\theta$ in momentum space, at different temperatures. Left (from \cite{Kanigel:2007}): underdoped Bi$_2$Sr$_2$CaCuO$_8$. Right (from \cite{Benini:2010qc}): the holographic model. In both a node is visible at low temperatures, while it opens up into an arc at higher temperatures.
\label{fig: gap}}
\end{figure}

Let us give some details on the resulting traced spectral function $A(\omega, \vec k) \equiv \frac1\pi \Tr \im G_R(\omega, \vec k)$, which can be computed numerically \cite{Benini:2010qc}. In figure \ref{fig: EDCs} we plot the energy distribution curves (EDCs), that is the spectral function computed along $\omega$ at the would-be Fermi momentum $k_F(\theta)$ and at different values of the angle $\theta$ in momentum space, at different temperatures and compare them with an experimental sample of underdoped Bi$_2$Sr$_2$CaCuO$_8$ \cite{Kanigel:2007}. From the EDCs we can extract the gap $\Delta_\omega(\theta)$ at different temperatures, plotted in fig. \ref{fig: gap} and compared with the sample. From the plot is evident the d-wave behavior $\Delta_\omega(\theta) \propto |\cos(2\theta)|$ with four nodes at low temperatures, and the development of four Fermi arcs at higher temperatures. In fig. \ref{fig: spectral function} a density plot of the spectral function in momentum space at $\omega=0$ also shows how Dirac nodes (left) open up into arcs (right). Despite this encouraging similarities, in the holographic model the arc length does not show a linear behavior with temperature.

\subsection{Electron stars}
\label{sec: electron stars}

Let us now come back to the simple bulk system of a graviton, a $U(1)$ gauge field and a Dirac fermion, as in \eqref{Lagrangian basic setup} and \eqref{Dirac Lagrangian}:
\be
\cL = \frac1{2\kappa^2} \Big( R + \frac6{L^2} \Big) - \frac1{4e^2} F_{\mu\nu} F^{\mu\nu} - i \overline{\psi} (\Gamma^\mu D_\mu - m) \psi \;.
\ee
This time, instead of treating the fermions as pure probes, one can notice that for $(mL)^2 \leq \gamma^2$ there is Schwinger pair production in the bulk \cite{Pioline:2005pf} which leads to the population of the Fermi sea \cite{Arsiwalla:2010bt, Hartnoll:2009ns, Hartnoll:2010gu, Hartnoll:2011dm}: the $U(1)$ is not broken (because Pauli exclusion principle prevents a macroscopic occupation of the ground state), rather a bulk Fermi surface appears. This represents a fermionic form of instability.

In the limit $mL, \gamma \gg 1$ one can use a WKB (or Thomas-Fermi-Oppenheimer-Volkov) approximation in which the Dirac eigenstates become very localized and the fermions can be treated as an ideal fluid \cite{Hartnoll:2009ns, Hartnoll:2010gu}:
\be
\cL = \frac1{2\kappa^2} \Big( R + \frac6{L^2} \Big) - \frac1{4e^2} F_{\mu\nu} F^{\mu\nu} + (\mu_\text{loc} \sigma - \rho) \;,
\ee
where $\mu_\text{loc}$ is the local bulk chemical potential, $\rho$ the energy density and $\sigma$ the charge density. Assuming that $T=0$ in the bulk, the equation of state is fixed by the following equations:
\bea
\mu_\text{loc} &= \frac{A_t}{\sqrt{g_{tt}}} \;,\qquad& p &= \mu_\text{loc} \sigma - \rho \;,\qquad \\
\rho &= \int_m^{\mu_\text{loc}} E \, g(E)\, dE \;,\qquad& \sigma &= \int_m^{\mu_\text{loc}} g(E)\, dE \;,\qquad& g(E) &= \frac E{\pi^2} \sqrt{E^2 - m^2} \;,
\eea
where $p$ is pressure and $g(E)$ is the density of states.
The whole system can be solved numerically.

\subsection{IR geometries and Lifshitz scaling}
\label{sec: Lifshitz}

In the presence of the bosonic (sec. \ref{sec: superconductors}) or fermionic (sec. \ref{sec: electron stars}) instabilities, the extra fields deform the core of the geometry. It turns out that in the IR a (non-relativistic) Lifshitz scaling symmetry
\be
r \to \lambda r \;,\qquad t \to \lambda^z t \;,\qquad (x,y) \to \lambda (x,y)
\ee
might emerge. Here $z$ is called ``dynamical exponent'': $z=1$ corresponds to a relativistic scaling, while $z=\infty$ is called ``local criticality''.

Indeed if we are only interested in the asymptotic IR geometry, for instance in the bosonic case \eqref{superconductor ansatz}, we can analytically look for solutions of the form:
\be
\frac{ds^2}{L^2} = - \frac{dt^2}{r^{2z}} + g_\infty \frac{dr^2}{r^2} + \frac{dx^2 + dy^2}{r^2} \;,\qquad A = \gamma\, h_\infty \frac{dt}{r^z} \;,\qquad \phi = \phi_\infty \;,
\ee
where $g_\infty, h_\infty, \phi_\infty$ are constants. The existence of such solutions depends on the choice of mass $m^2$ and potential $V$ in \eqref{Lagrangian superconductor}. In the fermionic case the background is the same, with $p = p_\infty$ and $\rho = \rho_\infty$. In this case for $e^2\gamma^2 \to \infty$ one finds $z \to 1$ and the geometry is AdS$_4$; for $e^2\gamma^2 \to 0$ one finds $z \to \infty$ and the geometry is $AdS_2 \times \bR^2$.

Let us remark some properties of the Lifshitz background. First, it does not have horizon: the electric flux---that was emanating from the black hole horizon in \eqref{RNAdS BH} before taking into account the extra unstable fields---now emanates from the boson or the fermion. Second, Lifshitz scaling implies that the temperature dependence of the entropy is $S \propto T^{2/z}$. Indeed at $T=0$ the entropy vanishes, while for $z \to \infty$ one finds a finite entropy ground state. Third, the Lifshitz geometry is not geodesically complete, and the question remains whether there is a production of excited string states in the far IR, leading to a further instability.

\section{Local criticality and the impurity problem}
\label{sec: impurity}

In section \ref{sec: AdS2} we saw that a simple holographic model can realize, as the $AdS_2 \times \bR^2$ background, a local quantum critical ground state, which is particularly interesting because of its relation to the strange metal phase. In fact it is also possible to realize such state with an impurity problem \cite{Sachdev:2010um, Sachdev:2010uj}.

For instance, following \cite{Sachdev:2010uj} consider a single spin impurity at $\vec x = 0$ coupled to the CFT$_3$ at the N\'eel/VBS antiferromagnetic transition discussed in sec. \ref{sec: emergent gauge}. Let us couple them through the following partition function:
\bea
Z &= \int \cD z^\alpha(x,\tau) \, \cD A_\mu(x,\tau) \, \cD\chi(\tau)\, \exp \Big\{ - \int d\tau\, \cL_\text{imp} - \int d^2x\, d\tau\, \cL_{z,A} \Big\} \\
\cL_\text{imp} &= i \chi^\dag  \Big( \parfrac{}{\tau} - iA_\tau(0,\tau) \Big) \chi \;,
\eea
where $\cL_{z,A}$ is the effective Lagrangian \eqref{Lag emergent gauge} written in terms of the slave fermion $z_\alpha$, and the impurity $\hat S^a$ has been written in terms of a slave fermion $\chi$ as well: $\hat S^a = \frac12 \chi^*_\alpha \sigma^a_{\alpha\beta} \chi_\beta$.
The impurity correlators can be computed at large $N$ for $N$-dimensional spins: at low temperature $T \ll \omega$  they decay with power-law in time:
\be
\langle \hat S^a(\tau) \hat S^b(0) \rangle \,\sim\, \delta^{ab} \Big| \frac{\pi T}{\sin(\pi T \tau)} \Big|^\gamma \,\xrightarrow{T \to 0}\, \delta^{ab} |\tau|^{-\gamma} \;.
\ee
Moreover the ground state has finite zero-temperature entropy. These are precisely the properties of the AdS$_2$ local quantum critical system, and after all local criticality is the expected scaling of an impurity.

Similar properties can be obtained in four dimensions by coupling a spin impurity to 4d $\cN=4$ SYM \cite{Kachru:2009xf, Mueck:2010ja}:
\bea
S &= \int d\tau\, \cL_\text{imp} + \int d^3x\, d\tau\, \cL_\text{SYM} \;, \\
\cL_\text{imp} &= \chi^{\dag a} \Big( \delta^b_a \parfrac{}{\tau} + i A_\tau(0,\tau)^b_a + i v^I \phi_I(0,\tau)^b_a \Big) \chi_b \;.
\eea
Indeed such system is very similar to the semi-holographic Fermi liquid of sec. \ref{sec: semiholographic}

\begin{acknowledgement}
I would like to thank the organizers of the ``XVII European Workshop on String Theory 2011'' (5-9 September, Padova, Italy) for the invitation to give this talk, and Rakibur Rahman and especially Chris Herzog and Amos Yarom for a very enjoyable collaboration and many interesting discussions and clarifications. My work is supported in part by the DOE grant DE-FG02-92ER40697.
\end{acknowledgement}


\bibliographystyle{fdp}
\bibliography{AdSCMT_benini}


%
%
%
%
%
%
%

\end{document}